\documentclass[12pt]{article}

\usepackage{amssymb}
\usepackage{epsfig}

\def\21{$SU(2) \otimes U(1) $}
\def\np#1#2#3{           {Nucl. Phys. }{\bf #1} (19#2) #3}
\def\pl#1#2#3{           {Phys. Lett. }{\bf #1} (19#2) #3}
\def\pr#1#2#3{           {Phys. Rev. }{\bf #1} (19#2) #3}

\def\prl#1#2#3{          {Phys. Rev. Lett. }{\bf #1} (19#2) #3}
\def\half{{\textstyle{1 \over 2}}} 
\def\s#1{\tilde{#1}}
\def\lsim{\raise0.3ex\hbox{$\;<$\kern-0.75em\raise-1.1ex\hbox{$\sim\;$}}}
\def\gsim{\raise0.3ex\hbox{$\;>$\kern-0.75em\raise-1.1ex\hbox{$\sim\;$}}}

\newcommand {\cost} {\cos \theta_{\tilde t}}
\newcommand{\wt}{\widetilde}
\newcommand {\chiz} [1] {\tilde{\chi}^{0}_{#1} }
\newcommand {\beq} {\begin{equation}}
\newcommand {\eeq} {\end{equation}}
\newcommand {\bea} {\begin{eqnarray}}
\newcommand {\eea} {\end{eqnarray}}
\newcommand{\mx}{\left[\begin{array}} 
\newcommand{\finmx}{\end{array}\right]} 
\def\nn{\nonumber}

\begin{document}

\title{
\begin{flushright} \small \rm
  hep-ph/0001033 \\ 
  FTUV/00-7 \\ 
  IFIC/00-07  \\
   LC-TH-2000-005
\end{flushright}
Light Stop: MSSM versus R--parity violation}
\author{W.~Porod, D.~Restrepo, and J.W.F.~Valle \\[0.5cm] \small
Inst.~de F\'\i sica Corpuscular (IFIC), CSIC - U. de Val\`encia, \\ \small
Edificio Institutos de Paterna, Apartado de Correos  2085\\ \small
E-46071--Val\`encia, Spain } 
\maketitle

\begin{abstract}
  We discuss the phenomenology of the lightest stops in models where
  R-parity is broken by bilinear terms. In this class of models we
  consider scenarios where the R-parity breaking two-body decay
  ${\tilde t}_1 \to \tau^+ \, b $ competes with the leading
  three-body decays ${\tilde t}_1 \to W^+ \, b
  \, {\tilde \chi}^0_1$, ${\tilde t}_1 \to H^+ \, b \, {\tilde
    \chi}^0_1$, ${\tilde t}_1 \to b \, {\tilde l}^+_i \, \nu_l$, and
  ${\tilde t}_1 \to b \, {\tilde \nu}_l \, l^+$ ($l =e,\mu,\tau$).  We
  demonstrate that the R--parity violating decay can be the dominant
  one.  In particular we focus on the implications for a future $e^+
  e^-$ Linear Collider.
\end{abstract}

\section{Introduction}

The search for supersymmetry (SUSY) \cite{susy,Haber84} plays an
important r\^ole in the experimental program at the colliders LEP2 and
Tevatron. It will be even more important at future colliders, e.g. an
upgraded Tevatron, LHC, an $e^+ e^-$ linear collider.  Therefore many
phenomenological studies have been carried out in recent years (see
e.g.~\cite{phen1,phen2,epem,tevatron} and references therein).  Most
of them have been carried out in the context of the minimal
supersymmetric standard model (MSSM) \cite{Haber84,Tata95}.  However,
neither gauge invariance nor supersymmetry requires the conservation
of R-parity. Indeed, there is considerable theoretical and
phenomenological interest in studying possible implications of
alternative scenarios~\cite{beyond} in which R-parity is
broken~\cite{aul,expl0,rpold,arca}.  The violation of R-parity could
arise explicitly~\cite{expl} as a residual effect of some larger
unified theory~\cite{expl0}, or spontaneously, through nonzero vacuum
expectation values (vev's) for scalar neutrinos~\cite{aul,rpold,arca}.
In realistic spontaneous R-parity breaking models there is an \21
singlet sneutrino vev characterizing the scale of R-parity violation
\cite{MASIpot3,MASI,ROMA,ZR} which is expected to be the same as the
effective supersymmetry breaking scale.

There are two generic cases of spontaneous R-parity breaking models to
consider.
In the absence of any additional gauge symmetry, these models lead to
the existence of a physical massless Nambu-Goldstone boson, called
majoron (J) which is {\sl the lightest SUSY particle}, massless and
therefore stable. 
If lepton number is part of the gauge symmetry and R-parity is
spontaneously broken then there is an additional gauge boson which
gets mass via the Higgs mechanism, and there is no physical Goldstone
boson~\cite{ZR}.  As in the standard case in R-parity breaking models the
lightest SUSY particle (LSP) is in general a neutralino. However, it
now decays mostly into visible states, therefore diluting the missing
momentum signal and bringing in increased multiplicity events which
arise mainly from three-body decays such as
\beq
\label{vis}
\chiz{1} \to  f \bar{f} \nu,
\eeq
where $f$ denotes a charged fermion. The neutralino also has the invisible
decay mode
\beq
\label{invi}
\chiz{1} \to  3 \nu.
\eeq
as well as
\beq
\chiz{1} \to  \nu  J ,
\label{invis}
\eeq
in the case the breaking of R-parity is spontaneous~\cite{MASIpot3,MASI}.
This last decay conserves R-parity since the majoron has a large R-odd
singlet sneutrino component.

Owing to the large top Yukawa coupling the stops have a quite
different phenomenology compared to those of the first two generations
of up--type squarks (see e.g.~\cite{Bartl97a} and references therein).
The large Yukawa coupling implies a large mixing between ${\tilde
  t}_L$ and ${\tilde t}_R$ \cite{Ellis83} and large couplings to the
higgsino components of neutralinos and charginos.  The large top quark
mass also implies the existence of scenarios where all two-body decay
modes of ${\tilde t}_1$ (e.g. ${\tilde t}_1 \to t \, {\tilde
  \chi}^0_i, b \, {\tilde \chi}^+_j, t \, \tilde g$) are kinematically
forbidden. In these scenarios higher order decays of ${\tilde t}_1$
become relevant: \cite{Hikasa87,Porod97}:
${\tilde t}_1 \to c \, {\tilde \chi}^0_{1,2}$, 
${\tilde t}_1 \to W^+ \, b \, {\tilde \chi}^0_1$,
${\tilde t}_1 \to H^+ \, b \, {\tilde \chi}^0_1$,
${\tilde t}_1 \to b \, {\tilde l}^+_i \, \nu_l$,
${\tilde t}_1 \to b \, {\tilde \nu}_l \, l^+$, 
%
where $l$ denotes $e,\mu,\tau$. In \cite{Porod97} it has been shown
that the three-body decay modes are in general much more important
than the two body decay mode in the framework of the MSSM.  Recently
it has been demonstrated that not only LSP decays are sign of R-parity
violation but that also the light stop is possible candidate for
observing R-parity violation even if R-parity violation is small
\cite{tevatron,stop1,stop2}. It has been demonstrated that there
exists a large parameter region where the R-parity violating decay
\beq
{\tilde t}_1 \to b \, \tau
\eeq
is much more important than
\beq
{\tilde t}_1 \to c \, {\tilde \chi}^0_{1,2} 
\eeq
in scenarios where only those decay modes are possible.  It is
therefore natural to ask if there exist scenarios where the decay
${\tilde t}_1 \to b \, \tau$ is as important as the three--body
decays.  Note that in the R-parity violating models under
consideration the neutral (charged) Higgs--bosons mix with the neutral
(charged) sleptons.  These states are denoted by $S^0_i$, $P^0_j$, and
$S^\pm_k$ for the neutral scalars, pseudoscalars and charged scalars,
respectively.  Therefore in the R-parity violating case one has the
following three-body decay modes:
\begin{eqnarray}
{\tilde t}_1 &\to& W^+ \, b \, {\tilde \chi}^0_1 \\
{\tilde t}_1 &\to& S^+_k \, b \, {\tilde \chi}^0_1 \\
{\tilde t}_1 &\to& S^+_k \, b \, \nu_l \\
{\tilde t}_1 &\to& b \, S^0_i \, l^+ \,,  \\
{\tilde t}_1 &\to& b \, P^0_j \, l^+ \,.
\end{eqnarray}
We will demonstrate that ${\tilde t}_1 \to b \, \tau^+$ can indeed be
the most important decay mode.  In particular we will consider a mass
range of ${\tilde t}_1$, where it is difficult for the LHC to discover
the light stop within the MSSM due to the large top background
\cite{ulrike}.  The rest of paper is organized in the following way:
in the next section we will introduce the model. In
Sect.~\ref{sec:num} numerical results for stop decays are presented
and their implications for LC.  In Sect.~4 we present our conclusions.

\section{The model}

The supersymmetric Lagrangian is specified by the superpotential $W$
given by
\begin{equation}  
W=\varepsilon_{ab}\left[ 
 h_U^{ij}\widehat Q_i^a\widehat U_j\widehat H_2^b 
+h_D^{ij}\widehat Q_i^b\widehat D_j\widehat H_1^a 
+h_E^{ij}\widehat L_i^b\widehat R_j\widehat H_1^a 
-\mu\widehat H_1^a\widehat H_2^b 
\right] + \varepsilon_{ab}\epsilon_i\widehat L_i^a\widehat H_2^b\,,
\label{eq:Wsuppot} 
\end{equation}
where $i,j=1,2,3$ are generation indices, $a,b=1,2$ are $SU(2)$
indices, and $\varepsilon$ is a completely antisymmetric $2\times2$
matrix, with $\varepsilon_{12}=1$. The symbol ``hat'' over each letter
indicates a superfield, with $\widehat Q_i$, $\widehat L_i$, $\widehat
H_1$, and $\widehat H_2$ being $SU(2)$ doublets with hypercharges
$1/3$, $-1$, $-1$, and $1$ respectively, and $\widehat U$,
$\widehat D$, and $\widehat R$ being $SU(2)$ singlets with
hypercharges $-{\textstyle{4\over 3}}$, ${\textstyle{2\over 3}}$, and
$2$ respectively. The couplings $h_U$, $h_D$ and $h_E$ are $3\times 3$
Yukawa matrices, and $\mu$ and $\epsilon_i$ are parameters with units
of mass.
 
Supersymmetry breaking is parametrized by the standard set of soft
supersymmetry breaking terms 
\begin{eqnarray} 
V_{soft}&=& 
M_Q^{ij2}\widetilde Q^{a*}_i\widetilde Q^a_j+M_U^{ij2} 
\widetilde U^*_i\widetilde U_j+M_D^{ij2}\widetilde D^*_i 
\widetilde D_j+M_L^{ij2}\widetilde L^{a*}_i\widetilde L^a_j+ 
M_R^{ij2}\widetilde R^*_i\widetilde R_j \nonumber\\ 
&&\!\!\!\!+m_{H_1}^2 H^{a*}_1 H^a_1+m_{H_2}^2 H^{a*}_2 H^a_2\nn\\
&&\!\!\!\!- \left[\half M_3\lambda_3\lambda_3+\half M\lambda_2\lambda_2 
+\half M'\lambda_1\lambda_1+h.c.\right] 
\nn\\ 
&&\!\!\!\!+\varepsilon_{ab}\left[ 
A_U^{ij}h_U^{ij}\widetilde Q_i^a\widetilde U_j H_2^b 
+A_D^{ij}h_D^{ij}\widetilde Q_i^b\widetilde D_j H_1^a 
+A_E^{ij}h_E^{ij}\widetilde L_i^b\widetilde R_j H_1^a\right.
\nn\\ 
&&\!\!\!\!\left.-B\mu H_1^a H_2^b+B_i\epsilon_i\widetilde L_i^a H_2^b\right] 
\,,\label{eq:Vsoft}
\end{eqnarray} 

Note that, in the presence of soft supersymmetry breaking terms the
bilinear terms $\epsilon_i$ can not be rotated away, since the rotation
that eliminates it reintroduces an R--Parity violating trilinear term,
as well as a sneutrino vacuum expectation value~\cite{BRpV}.

For our discussion it suffices to assume R-parity Violation (RPV) only
in the third generation. However we do allow for R-parity-conserving
Flavour Changing Neutral Currents (FCNC) effects, such as the process
$\s t_1\to c\,\s\chi_1^0$ involving the three generations of quarks.
In this case we will omit the labels $i,j$ in the soft breaking terms.
In order to study the R--Parity violating decay mode $\s t_1\to b\,
\tau$ it is sufficient to consider the superpotential
\cite{BRpV,moreBRpV,otros,javi}
\begin{equation} 
W=h_t\widehat Q_3\widehat U_3\widehat H_2
 +h_b\widehat Q_3\widehat D_3\widehat H_1
 +h_{\tau}\widehat L_3\widehat R_3\widehat H_1
 -\mu\widehat H_1\widehat H_2
 +\epsilon_3\widehat L_3\widehat H_2
\label{eq:Wbil}
\end{equation}
This amounts to neglecting the effects of RPV on the two first
families. A short discussion on ${\tilde t}_1 \to b \, l^+$ 
in the three generation model will be given at the end of
Sect.~\ref{sec:num}.

The bilinear term in Eq.~(\ref{eq:Wbil}) leads to a mixing between
the charginos and the $\tau$--lepton which in turn leads to the decay
${\tilde t}_1 \to b \, \tau$. The mass matrix is given by
\begin{equation}
{\bf M_C}=\left[\matrix{
M & {\textstyle{1\over{\sqrt{2}}}}gv_2 & 0 \cr
{\textstyle{1\over{\sqrt{2}}}}gv_d & \mu & 
-{\textstyle{1\over{\sqrt{2}}}}h_{\tau}v_3 \cr
{\textstyle{1\over{\sqrt{2}}}}gv_3 & -\epsilon_3 &
{\textstyle{1\over{\sqrt{2}}}}h_{\tau}v_d}\right]
\label{eq:ChaM6x6}
\end{equation}
As in the MSSM, the chargino mass matrix is diagonalized by two
rotation matrices $\bf U$ and $\bf V$
\begin{equation}
{\bf U}^*{\bf M_C}{\bf V}^{-1}=\left[\matrix{
m_{\s\chi^{\pm}_1} & 0 & 0 \cr
0 & m_{\s\chi^{\pm}_2} & 0 \cr
0 & 0 & m_{\tau}}\right]\,.
\label{eq:ChaM3x3}
\end{equation}
The lightest eigenstate of this mass matrix must be the tau lepton
($\tau^{\pm}$) and so the mass is constrained to be 1.7771 GeV.  To
obtain this the tau Yukawa coupling becomes a function of the
parameters in the mass matrix, and the full expression is given
in~\cite{v3cha}.

The stop mass matrix is given by
\beq
M_{\tilde t}^2=\left[ \begin{array}{cc}
{M_Q^2}+\frac12v_u^2 {h_t}^2+\Delta_{UL}&
\frac{h_U}{\sqrt2}
  \left( v_u{A_t}- \mu v_d +\epsilon_3 v_3 \right) \\
\frac{h_U}{\sqrt2}
  \left( v_u{A_t}- \mu v_d +\epsilon_3 v_3 \right) &
{M_U^2}+\frac12 v_u^2{h_t}^2+\Delta_{UR}
\end{array} \right]
\eeq
with $\Delta_{UL}=\frac18\big(g^2-\frac13{g'}^2\big)\big(v_d^2-v_u^2
+v_3^2\big)$ and $\Delta_{UR}=\frac16 {g'}^2(v_d^2-v_u^2+v_3^2)$.
The sum of the $v^2_i$ is given by $m^2_W = g^2 (v_d^2+v_u^2+v_3^2)/2$.
The mass eigenstates are given by
$
\wt t_1 = \wt t_L \cos \theta_{\wt t} + 
 \wt t_R \sin \theta_{\wt t}$ and 
$\wt t_2 = \wt t_R \cos \theta_{\wt t} - 
 \wt t_L \sin \theta_{\wt t}.
$
The sfermion mixing angle is given by
\beq
\cos \theta_{\wt t} = \frac{- M^2_{\wt t_{12}}}{\sqrt{(M^2_{\wt t_{11}} -
m^2_{\wt t_1})^2 + (M^2_{\wt t_{12}})^2}}, \qquad 
\sin \theta_{\wt t} = \frac{M^2_{\wt t_{11}} - m^2_{\wt t_1}}
{\sqrt{(M^2_{\wt t_{11}} - m^2_{\wt t_1})^2 + (M^2_{\wt t_{12}})^2}} \, .
\eeq

In addition the charged Higgs bosons mix with charged sleptons and the real 
(imaginary) parts of the sneutrino mix the scalar (pseudoscalar) Higgs
bosons. The formulas can be found e.g. in \cite{v3cha,stop3}.
Their main decay modes for the mass range considered in this study are:
\bea
S^0_i &\to & b \, \bar{b} \, , \, \tau^+ \tau^- \, , \, \chiz{1} \nu_\tau \\
P^0_j &\to & b \, \bar{b} \, , \, \tau^+ \tau^- \, , \, \chiz{1} \nu_\tau \\
S^-_k &\to & s \, \bar{c} \, , \, \tau^- \nu_\tau \, , \, \chiz{1} \tau^-
\eea

\section{Numerical results}
\label{sec:num}

In this section we present our numerical results for the branching
ratios of the higher order decays of ${\tilde t}_1$. Here we consider
scenarios where all two-body decays induced at tree-level are
kinematically forbidden.  Before going into detail it is useful to
have some approximate formulas at hand \cite{stop2}:
\bea
\Gamma(\s t_1\to b\,\tau) &\approx&
\frac{g^2 |U_{32}|^2 h_b^2 \cos^2_{\theta_{\s t}} m_{\s t_1}}{16\pi} 
\label{tausa} \\
\Gamma(\s t_1\to c\,\s\chi_1^0)&\approx& 10^{-6}h_b^4m_{\s t_1} \eea
where $|U_{32}| \approx | \epsilon_3 / \mu|$ if $|\epsilon_3| \ll
|\mu|$ and $v_3 \ll m_W$. The complete formulas are given in
\cite{stop1,stop2}.  For the three--body decays the formulas given in
\cite{Porod97} can be used as a good approximation if the mixings
induced by R-parity violation are small. The complete formulas for the
three--body decays in the R--parity violating case will be given
elsewhere \cite{stop3}.

We have fixed the parameters as in \cite{Porod97} to avoid colour
breaking minima: we have used $m_{{\tilde t}_1}$, $\cos \theta_{\tilde
  t}$, $\tan \beta$, and $\mu$ as input parameters in the top squark
sector.  For the sbottom (stau) sector we have fixed $M_{\tilde Q},
M_{\tilde D}$ and $A_b$ ($M_{\tilde E}, M_{\tilde L}$, and $A_\tau$)
as input parameters.  In addition we choose the R-parity violating
parameters $\epsilon_3$ and $v_3$ in such a way that the tau neutrino
mass is fixed (\cite{stop2} and references therein):
\begin{equation}
m_{\nu_{\tau}} \approx 
-\frac{(g^2M'+g'^2M){\mu'}^2}{
4 M M'{\mu'}^2-2(g^2M'+g'^2M){\mu'}v_u{v_d'} \cos\xi}{v_d'}^2\sin^2\xi
\label{mNeutrinoApp}
\end{equation}
with
\bea
\sin\xi &=& \frac{\epsilon_3 v_d +\mu  v_3 }
               {\sqrt{\mu^2+\epsilon_3^2} \sqrt{v_d^2+v_3^2}} \\
\mu' &=& \sqrt{\mu^2+\epsilon_3^2} \, , \hspace{2cm}
v'_d = \sqrt{v_d^2+v_3^2} \,.
\eea
For simplicity, we assume that the soft SUSY breaking parameters are
equal for all generations. 
\begin{figure}
\setlength{\unitlength}{1mm}
\begin{picture}(150,50)
\put(-4,-4){\mbox{\epsfig{figure=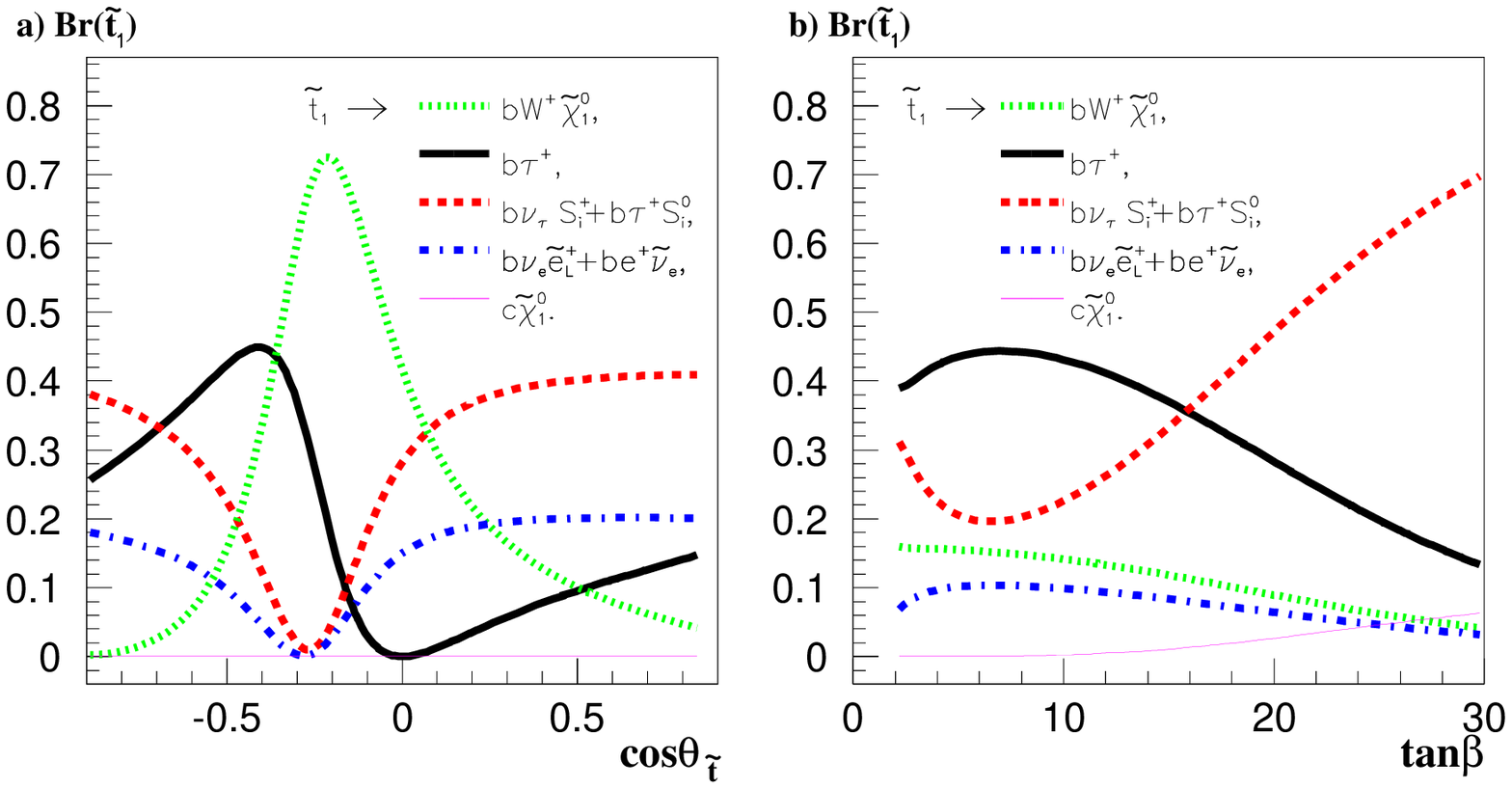,height=7.0cm,width=15.6cm}}}
\end{picture}
\caption[]{Branching ratios for ${\tilde t}_1$ decays
  for $m_{{\tilde t}_1} = 220$~GeV, $\mu = 500$~GeV, $M =
  240$~GeV, and $m_\nu = 100$~eV.
  In a) the branching ratios are shown as a function of
  $\cos \theta_{\tilde t}$ for $\tan \beta = 4$, in b) as a function
  of $\tan \beta$ for $\cost = 0.25$. Note, that the graph - - displays
  the sum
  $b \, \nu_\tau \, S^+_i + b \, \tau^+ \, S^0_i +b \, \tau^+ \, P^0_i$.}
\label{brst3cosa}
\end{figure}

\begin{table}
\begin{tabular}{|l|lll|}\hline
Input: & $\tan \beta = 4$ & $\mu = 500$ GeV & $M = 240$ GeV \\
 & $M_{\tilde D}=370$ GeV & $M_{\tilde Q}=340$ GeV & $A_b=150$ GeV \\
& $M_{\tilde E}=190$ GeV & $M_{\tilde L}=190$ GeV & $A_\tau=150$ GeV \\ 
& $m_{{\tilde t}_1}=220$ GeV & $\cos \theta_{\tilde t}=0.25$ &
 $m_{P^0_2}=110$ GeV \\ \hline
Calculated & $m_{{\tilde \chi}^0_1}=120$ &
       $m_{{\tilde \chi}^+_1}=225$ & $m_{{\tilde \chi}^+_2}=520$  \\
 &  $m_{{\tilde b}_1}=340$ GeV & $m_{{\tilde b}_2}=375$ GeV & 
      $\cos \theta_{\tilde b}=0.925$ \\
  & $m_{S^0_1}=82$ GeV & $m_{S^0_2}=128$ GeV & $m_{S^0_3}=182$ GeV \\
  &  $m_{P^0_2}=110$ GeV & $m_{P^0_3}=182$ GeV  & \\
  & $m_{S^-_2}=136$ GeV & $m_{S^-_3}=187$ GeV & $m_{S^-_4}=204$ GeV  \\
 & $m_{{\tilde e}_L}=213$ GeV &
      $m_{{\tilde \nu}_e}=m_{{\tilde \nu}_\tau}=204$ GeV  & \\ \hline
\end{tabular}
\caption[]{Input parameters and resulting quantities used in Fig.~1.}
\label{tabbrst3cosa}
\end{table}

In Fig.~\ref{brst3cosa}(a) and (b) we show the branching ratios of
${\tilde t}_1$ as a function of $\cos \theta_{\tilde t}$.  The
parameters and physical quantities are given in Tab.~\ref{tabbrst3cosa}.
In Fig.~\ref{brst3cosa}(a) we show $\mbox{BR}({\tilde t}_1 \to b \,
\tau^+)$, $\mbox{BR}({\tilde t}_1 \to c \, {\tilde \chi}^0_1)$,
$\mbox{BR}({\tilde t}_1 \to b \, W^+ \, {\tilde \chi}^0_1)$,
$\mbox{BR}({\tilde t}_1 \to b \, e^+ \, \tilde{\nu}_e$) +
$\mbox{BR}({\tilde t}_1 \to b \, \nu_e \, \tilde{e}^+_L)$.
The
branching ratios for decays into $\tilde{\mu}_{L}$ or
$\tilde{\nu}_{\mu}$ are practically the same as those into
$\tilde{e}_{L}$ or $\tilde{\nu}_{e}$. 
We have summed up those branching ratios for the decays into sleptons
that give the same final state, for example:
\begin{eqnarray}
{\tilde t}_1 \to b \, \nu_e \, {\tilde e}^+_L \,
          \to \, b \, e^+ \, \nu_e \, {\tilde \chi}^0_1 \;, \hspace{5mm}
{\tilde t}_1 \to b \, e^+ \, {\tilde \nu}_e \,
          \to \, b \, e^+ \, \nu_e \, {\tilde \chi}^0_1
\end{eqnarray}
Note that in Fig.~\ref{brst3cosa} we have also summed the decay
branching ratios $\mbox{BR}({\tilde t}_1 \to b \, S^+_k \, \nu_\tau)$
+ $\mbox{BR}({\tilde t}_1 \to b \, \tau^+ \, S^0_i)$ +
$\mbox{BR}({\tilde t}_1 \to b \, \tau^+ \, P^0_j)$.

In the above cases the assumption $m_{{\tilde t}_1} - m_b <
m_{{\tilde\chi}^+_1}$ implies $m_{{\tilde\chi}^+_1} > m_{\tilde l}$.
Therefore, charginos can not arise as decay products of sleptons. The
latter can only decay into the corresponding lepton plus ${\tilde
  \chi}^0_1$ except for a small parameter region where the decay into
${\tilde \chi}^0_2$ is possible.  However, this decay is negligible
due to kinematics in that region.  In addition there exists the
possibility of R-parity violating decays. However, these will be small
because the neutrinos mix mainly with higgsinos implying that the
partial decay widths are proportional to the squared product of an
R-parity violating mixing parameter and small Yukawa coupling.  
 For this set of parameters
$\mbox{BR}({\tilde t}_1 \to c \, {\tilde \chi}^0_1)$ is $O(10^{-4})$
independent of $\cos \theta_{\tilde t}$ and therefore negligible. Near
$\cos \theta_{\tilde t} = -0.3$ one has ${\tilde t}_1 \to b \, W^+ \,
{\tilde \chi}^0_1$ as dominant decay channel, since the ${\tilde
  t}_1$-${\tilde \chi}^+_1$-$b$ coupling vanishes implying that the
main contribution for the decays into the scalars vanishes.  Moreover,
the width for ${\tilde t}_1 \to b \, \tau^+$ is somewhat suppressed
because of the $\cos^2 \theta_{\tilde t}$ factor in Eq.~(\ref{tausa}).
Note, from the figure that the branching ratios for the various decays
into selectrons $e$-sneutrino is roughly a factor two smaller than the
sum of the decays into $S^\pm_k$, $S^0_i$, $P^0_j$.  The reason is
that, for this choice of parameters the $ P^0_2$ is mainly the pseudoscalar
Higgs boson $A^0$ of the MSSM with mass 110 GeV. In this case the R-parity
violating channel ${\tilde t}_1 \to b \, \tau^+ \, P^0_2$ is
comparable to the corresponding R-parity conserving decays.  This
state appears additional to the states which carry tau--lepton number
in the MSSM limit giving rise to the observed difference. Note, that
one has to expect additional jets from the states containing the
scalars $S^0_i$, $P^0_j$, and $S^\pm_k$ because they have admixtures
of the original Higgs boson.  In case of negative $\cost$ the decay
into ${\tilde t}_1 \to b \, \tau^+$ is important and can be even the
most important one.  Therefore one has events with $\tau^+ \, \tau^-
\, b \, \bar{b}$ in the final state which can be used for a full mass
reconstruction of the light stop.

In Fig.~\ref{brst3cosa}(b) the $\tan \beta$ dependence of branching
ratios is shown. For this specific choice of $\cost$ the decay
${\tilde t}_1 \to b \, \tau^+$ is the most important one for $\tan
\beta \lsim 15$. Above this value the final states which contain the
scalars corresponding to the lighter MSSM stau are the most important
ones.  The growth of the decay branching ratios into these states with
$\tan \beta$ is a feature independent of $\cost$.

\begin{figure}
\setlength{\unitlength}{1mm}
\begin{picture}(150,50)
\put(-4,-4){\mbox{\epsfig{figure=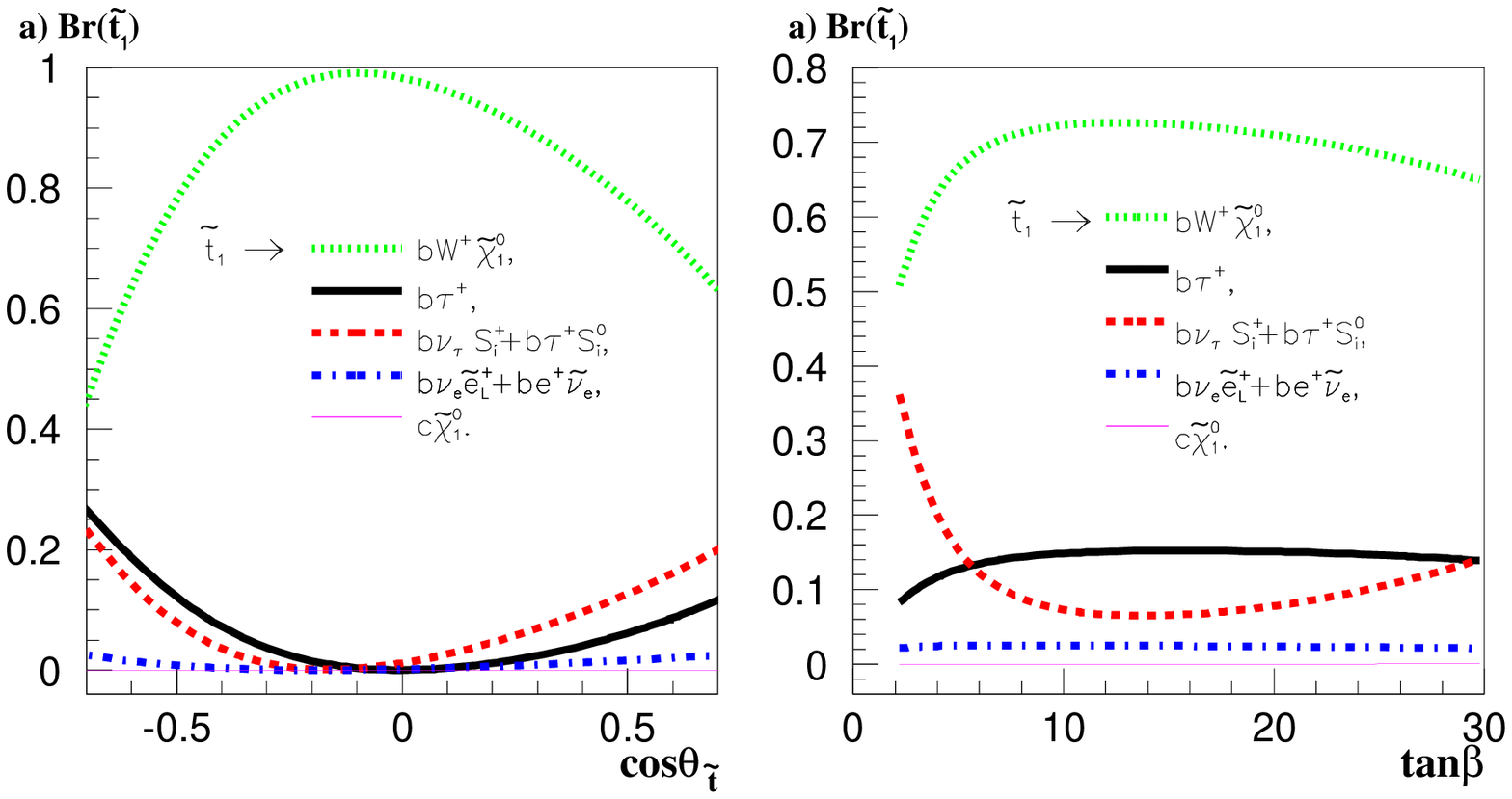,height=7.0cm,width=15.6cm}}}
\end{picture}
\caption[]{Branching ratios for ${\tilde t}_1$ decays 
  for $m_{{\tilde t}_1} = 350$~GeV, $\mu = 750$~GeV, $M =
  380$~GeV, and $m_\nu = 1$~keV.
 In a) the branching ratios are shown as a function of $\cos
  \theta_{\tilde t}$ for $\tan \beta = 4$, in b) as a function of
  $\tan \beta$ for $\cost = 0.7$. Note, that the graph - - displays
  the sum
  $b \, \nu_\tau \, S^+_i + b \, \tau^+ \, S^0_i +b \, \tau^+ \, P^0_i$.}
\label{brst3cosb}
\end{figure}

The assumption that no tree-level-induced two-body decays are
kinematically allowed implies that $m_{{\tilde\chi}^+_1} > m_{{\tilde
    t}_1} - m_b$.  Therefore, one expects an increase of
$\mbox{BR}({\tilde t}_1 \to b \, W^+ \, {\tilde \chi}^0_1)$ if
$m_{{\tilde t}_1}$ increases, because the decay into $b \, W^+ \,
{\tilde \chi}^0_1$ is dominated by the $t$ exchange, whereas for the
decays into scalars ${\tilde \chi}^+_j$ exchange dominates.  This
trend is indeed observed in Fig.~\ref{brst3cosb}, where we show the
branching ratios for $m_{{\tilde t}_1} = 350$~GeV. Here we concentrate
on the range of $\cost$ where $A_t \le 1$~TeV to avoid possible minima
in the scalar potential which break either color or electric charge.
Notice that for the heavy stop case the decay $ {\tilde t}_1 \to b \,
W^+ \, {\tilde \chi}^0_1$ is the most important one, independently of
$\cost$ and $\tan \beta$.  

Note however that also in this case R-parity violation implies a
distinct signature compared to what is expected in the MSSM due to the
decays of ${\tilde \chi}^0_1$. One gets the following
high-multiplicity final states:
\bea
{\tilde t}_1 &\to& b \, W^+ \, f \, \bar{f} \, \nu  \\
&\to & b \, W^+ \, f \, \bar{f}' \, l^\pm \\
&\to& b \, W^+ \, \nu \, J 
\eea 
where $f$ denotes a Standard Model fermion. Here the decays of the
$W$--boson will give additional leptons and jets.  Therefore, one has
in general additional jets and leptons compared to the MSSM case.
\begin{figure}
\setlength{\unitlength}{1mm}
\begin{picture}(150,60)
\put(-9,-4){\mbox{\epsfig{figure=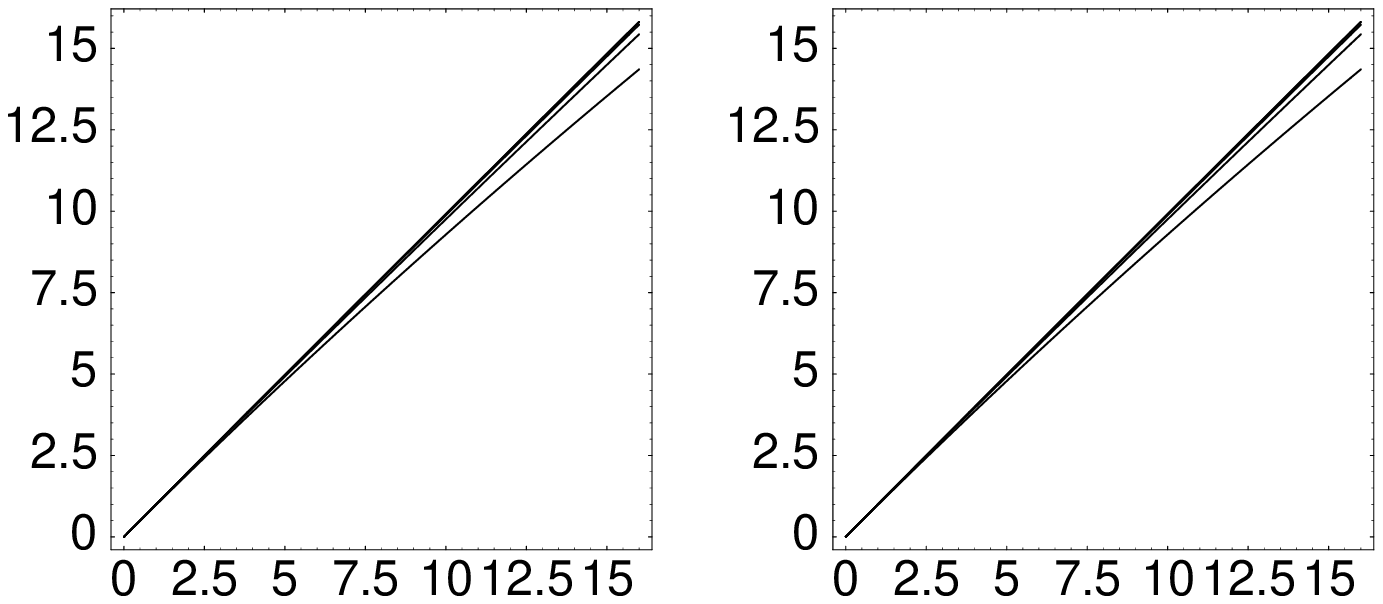,height=6.0cm,width=15.cm}}}
\put(-1,55){{\small \bf a)}}
\put(5,53){\makebox(0,0)[bl]{{\small $\mbox{BR}({\tilde t}_1 \to b \mu)/
                                        \mbox{BR}({\tilde t}_1 \to b \tau) $}}}
\put(64,-3){\makebox(0,0)[br]{{\small $(\epsilon_2 / \epsilon_3)^2$}}}
\put(29,41){{\small $\cost = 0.8$}}
\put(39,29){{\small $\cost = 0.05$}}
\put(69,55){{\small \bf b)}}
\put(75,53){\makebox(0,0)[bl]{{\small $\mbox{BR}({\tilde t}_1 \to b e)/
                                        \mbox{BR}({\tilde t}_1 \to b \tau) $}}}
\put(137,-3){\makebox(0,0)[br]{{\small $(\epsilon_1 / \epsilon_3)^2$}}}
\put(102.5,41){{\small $\cost = 0.8$}}
\put(112.5,29){{\small $\cost = 0.05$}}
\end{picture}
\caption[]{Ratio of branching ratios: a)$\mbox{BR}({\tilde t}_1 \to b \mu)/
  \mbox{BR}({\tilde t}_1 \to b \tau) $ as a function of $(\epsilon_2 /
  \epsilon_3)^2$ and b) $\mbox{BR}({\tilde t}_1 \to b e)/
  \mbox{BR}({\tilde t}_1 \to b \tau) $ as a function of $(\epsilon_1 /
  \epsilon_3)^2$ for for $m_{{\tilde t}_1} = 220$~GeV, $\mu =
  500$~GeV, $M = 240$~GeV, and $\epsilon_1^2 + \epsilon_2^2 +
  \epsilon^2_3 = 1$~GeV$^2$; $\cost = 0.05, 0.1, 0.2, 0.4 0.8$,
  $m_{\nu_\tau} = 100$~eV. }
\label{breps}
\end{figure}

In the event that $\epsilon_{1,2}$ are of the same order of magnitude,
as suggested by a solution to the present neutrino
anomalies~\cite{romao} one has in addition the decays into $ b \, e^+$
and $ b \, \mu^+$.  If one passes from the 1-generation model to the
3-generation model the situation changes as follows. From
Eq.~(\ref{tausa}) it follows that the sum of the modes $\Gamma({\tilde
  t}_1 \to b \, l^+)$ in the 3--generation model is nearly equal to
$\Gamma({\tilde t}_1 \to b \, \tau^+)$ in the 1--generation model, if
$(\epsilon'_1)^2 + (\epsilon'_2)^2 + (\epsilon'_2)^2 = \epsilon_3^2$
where the $\epsilon'_i$ ($\epsilon_3$) are the parameters of the
3--generation (1--generation) model.
In Fig.~\ref{breps} we show the ratios of branching ratios for
different $ {\tilde t}_1 \to b \, l^+$ modes versus the ratios of
different $\epsilon$'s squared and for different values of $\cost$. In
both cases we have fixed $\epsilon_1^2 + \epsilon_2^2 + \epsilon^2_3 =
1$~GeV$^2$ and in Fig.~\ref{breps}a $\epsilon_1 = \epsilon_3$ whereas
in Fig.~\ref{breps}b $\epsilon_2 = \epsilon_3$.  One can see that the
dependence is nearly linear even for rather small $\cost$. This result
depends on $(\epsilon_1^2 + \epsilon_2^2 + \epsilon^2_3)/\mu^2$ and on
the neutrino mass $m_{\nu_\tau}$, since both determine the mixings of
the leptons with the charginos.  The lines indicated in the figure
come closer to the diagonal if $(\epsilon_1^2 + \epsilon_2^2 +
\epsilon^2_3)/\mu^2$ increases and $m_{\nu_\tau}$ decreases.

\section{Conclusions}

We have studied the phenomenology of the lightest stop in scenarios
where the R-parity violating decay ${\tilde t}_1 \to b \, \tau^+$
competes with three--body decays. We have found that for $m_{{\tilde
    t}_1} \lsim 250$~GeV there are regions of parameter where ${\tilde
  t}_1 \to b \, \tau^+$ is an important decay mode if not the most
important one. This implies that there exists the possibility for full
stop mass reconstruction from $\tau^+ \, \tau^- \, b \, \bar{b}$ final
states. Moreover, in this mass range the discovery of the lightest stop
might not be possible at the LHC (certainly this is the case in the
MSSM).  This implies that one has to take into account the importance
of this new decay mode when designing the stop search strategies at a
future $e^+ e^-$ Linear Collider.  Spontaneously and bilinearly broken
R-parity violation imply additional leptons and/or jets in stop
cascade decays.  Looking at the three generation model the decays into
${\tilde t}_1 \to b \, l^+$ imply the possibility of measuring
$\epsilon^2_e /\epsilon^2_\tau$ and $\epsilon^2_\mu /\epsilon^2_\tau$
and thereby probing the parameters associated with the present solar
and atmospheric neutrino anomalies.

\section*{Acknowledgments}

This work was supported by DGICYT under grants PB95-1077 and by the
TMR network grant ERBFMRXCT960090 of the European Union. D.R.~was
supported by Colombian COLCIENCIAS fellowship.  W.P.~was supported by
the Spanish 'Ministerio de Educacion y Cultura' under the contract
SB97-BU0475382.


\end{document}